\documentclass{ws-mpla}
\usepackage[super]{cite}
\usepackage{graphicx}
\usepackage{enumerate}
\usepackage{subfigure}
\usepackage{color}
\usepackage{longtable}
\usepackage{multirow}

\begin{document}

\markboth{Lu Wang and Wenjie Liu}{A quantum segmentation algorithm based on local adaptive threshold
for NEQR image}

%%%%%%%%%%%%%%%%%%%%% Publisher's Area please ignore %%%%%%%%%%%%%%
\catchline{}{}{}{}{}
%%%%%%%%%%%%%%%%%%%%%%%%%%%%%%%%%%%%%%%%%%%%%%%%%%%%%%%%%%%%%%%%%%%

\title{A  quantum segmentation algorithm based on  local adaptive threshold  for NEQR image}

\author{Lu Wang}

\address{School of Automation, Nanjing University of Information Science  and Technology, \\
Nanjing, 210044, Jiangsu, China\\
Lu\_Wang\_MT@163.com}

\author{Wenjie Liu*}

\address{School of Computer and Software, Nanjing University of Information Science  and Technology, \\
Nanjing, 210044, Jiangsu, China\\
Engineering Research Center of Digital Forensics, Ministry of Education, Nanjing University of Information Science and Technology,\\
Nanjing, 210044, Jiangsu, China\\
wenjiel@163.com}

\maketitle

\pub{Received (Day Month Year)}{Revised (Day Month Year)}

\begin{abstract}
The classical image segmentation algorithm based on local adaptive threshold can effectively segment  images with uneven illumination, but with the increase of the  image data, the real-time problem gradually emerges. In this paper, a quantum segmentation algorithm based on local adaptive threshold for NEQR image is proposed, which can use quantum mechanism to simultaneously compute local thresholds for all pixels in a gray-scale image and quickly segment the image into a binary image.  In addition, several quantum circuit units, including median calculation, quantum binarization, etc., are designed in detail, and then  a complete quantum  circuit is designed to segment NEQR images by using fewer qubits and quantum gates. For a $2^n \times 2^n$ image with $q$ gray-scale levels, the complexity of our algorithm can be reduced to O$(n^2+q)$, which is an exponential speedup compared to the classic counterparts. Finally, the experiment is conducted on IBM Q to show the feasibility of our algorithm in the noisy intermediate-scale quantum (NISQ) era.

\end{abstract}

\ccode{PACS No.: 03.67.-a}

\section{Introduction}\label{sec1}

With the increase of our requirements for image quality, the amount of image data also increases rapidly, which requires a lot of computing power to process image. However, the computing power of classical computers is now approaching the limit, so the real-time problem is gradually revealed. Quantum image processing (QIP) is an emerging interdisciplinary subject that combines quantum computing and classical image processing, and has received extensive attention and research by researchers in recent years. It uses the unique superposition and parallelism characteristics of quantum computing to quickly improve the computing speed. Compared with classical image processing, QIP algorithms can achieve an exponential improvement \cite{Yan2017,Cai2018}.

The main work of QIP starts with how to represent images in a quantum computer. Some researchers have proposed some QIP representation models by encoding image information into probability magnitudes of quantum states, such as qubit lattice representation (QLR) \cite{Venegas-Andraca2003}, real ket representation (RKR) \cite{Latorre2005}, the flexible representation of quantum image (FRQI) \cite{Le2011}, the multi-channel RGB images representation of quantum images (MCQI) \cite{Sun2013}, a normal arbitrary superposition state of quantum image (NASS) \cite{Li2014} and quantum probability image encoding representation (QPIE) \cite{Yao2018}. This encoding method can efficiently encode images using fewer qubits, but it also makes it more difficult to retrieve images. To solve this problem, a novel enhanced quantum representation (NEQR) \cite{Zhang2013} is proposed, which is an encoding method that uses 
some sequences of qubits to encode images, which encodes the  position information and the gray-scale value information into three entangled qubit sequences, such that images can be retrieved quickly with few measurements. Therefore, the NEQR model is widely used due
to its simplicity of operation. By improving the position information and color information, a series of encoding methods have been proposed, such as an improved NEQR (INEQR) \cite{Jiang2015}, a generalized model of NEQR (GNEQR) \cite{Zhang2015L}, and a novel quantum representation of color digital images (NCQI ) \cite{Li2018}. Besides image coding based on probability magnitude and qubit sequence, there are some new techniques applied to image coding, such as a quantum log-polar image (QUALPI)  \cite{ZhangY2013} for processing log-polar sampled images, a Fourier transform qubit representation (FTQR) \cite{Grigoryan2020} for encoding color information into a quantum superposition state, and a bit-plane representation for log-polar quantum images (BRLQI) \cite{Chen2022}. With the development of quantum image representation models, corresponding quantum image processing algorithms have also developed rapidly, such as geometrical transformation of quantum image \cite{Le2010,Le2011S,Fan2016,Wang2014},quantum image steganography based on least significant bit (LSB) \cite{Jiang2016}, feature extraction of quantum image \cite{Hancock2015}, quantum image scrambling
\cite{Jiang2014}, quantum image watermarking \cite{Mogos2009,Iliyasu2012,Song2014,Miyake2016,Shahrokh2015}, 
quantum image filtering \cite{Li2017}, quantum image scaling \cite{Sang2016}, 
quantum image matching \cite{Jiang2016QIM}, 
quantum image edge detection \cite{Zhang2015,Fan2019,Zhou2019,Chetia2021}, 
quantum image segmentation \cite{Li2013,Caraiman2014,Caraiman2015,Xia2019,Yuan2020}, etc. However, most of the current QIP algorithms are only theoretically feasible because they use too many quantum resource, which is inappropriate in this noisy intermediate-scale quantum(NISQ) era.

Quantum image segmentation algorithm is the basis of image processing, and in recent years, it have gradually developed from theoretically feasible to practically feasible. In 2013, Li et al. \cite{Li2013} used
a quantum search algorithm to segment the quantum image,  but no specific oracle was provided.  In 2014, Caraiman et al. \cite{Caraiman2014} proposed a histogram-based  quantum image segmentation algorithm, and they segmented the image according to the distribution of pixels’ gray-scale values in the histogram and obtained exponential acceleration, but there was no specific oracle implementation circuit. One year later, they \cite{Caraiman2015} proposed a quantum image segmentation algorithm based on a single threshold and gave an oracle circuit, but they could not be simulated in the noisy intermediate-scale quantum (NISQ) era due to the high quantum cost. In 2019, Xia et al. \cite{Xia2019} designed a quantum comparator and used a single threshold to binarize the quantum image. They designed a detailed segmentation circuit but the complexity is still too high. Although many researchers have conducted research on
quantum image segmentation algorithms, but none of the previous algorithms are simulated on the quantum platform. In 2020, Yuan et al. \cite{Yuan2020} proposed a dual-threshold quantum image segmentation algorithm based on the quantum comparator they designed,  which is feasible in this NISQ era due to its low quantum cost. In addition, the algorithm is simulated  on the IBM Q platform \cite{IBM} to demonstrate its feasibility. However, like all the above algorithms, it can only use global fixed thresholds to segment the image, which is not effective for complex image (images with uneven illumination) segmentation,  Therefore, in order to better segment the quantum image and adapt to the current quantum situation, we conducted in-depth research in this paper, and the main contributions are as follows.
\begin{itemize}
    \item A quantum segmentation algorithm based on local adaptive threshold  for NEQR image is proposed, which can  accurately segment the image with uneven illumination into a binary image (the existing quantum segmentation algorithms cannot), and can solve the real-time problem of the classical counterpart.
    \item Several specific quantum circuit units are individually designed in detail, and then, based on these  units, a complete quantum image segmentation circuit is designed to segment the NEQR image  by using fewer qubits and quantum gates.
    \item  By analyzing  the  circuit  quantum cost and the MSE value of result image, we  verify  the  superiority    of  our  proposed  algorithm. Then,  the    the simulation  experiment is performed on  IBM Q platform \cite{IBM} to show the feasibility of our algorithm.
\end{itemize}

The rest of this paper is organized as follows: In Sect. \ref{sec2}, we  briefly introduce the  basic  quantum gates, the  NEQR  model and the classical local adaptive threshold for image
segmentation. In Sect. \ref{sec3},   several quantum operations is introduced firstly, and then, the quantum segmentation algorithm is proposed. Next, several specific quantum circuit units are  designed in detail, and a complete quantum image segmentation circuit is designed. In Sect. \ref{sec4}, the circuit complexity is analyzed, and the simulation experiment is conducted on the IBM Q platform to show the feasibility of our algorithm. Finally, the conclusion is given in Sect. \ref{sec5}.

\section{Preliminaries}\label{sec2}

\subsection{Quantum gates}

The single qubit states $\lvert 0 \rangle $ and $\lvert 1 \rangle $ are defined as follows,
\begin{equation}
    \lvert 0 \rangle {\rm{ = }}\left[ {\begin{array}{*{20}{c}}
1\\
0
\end{array}} \right],\lvert 1 \rangle {\rm{ = }}\left[ {\begin{array}{*{20}{c}}
0\\
1
\end{array}} \right].
\end{equation}

Unitary matrices of Hadamard (H), NOT gate (X), Controlled-NOT gate (CNOT), Toffoli gate (CCNOT) and Fredkin gate (CSWAP) can be expressed
as follows.
\begin{equation}
\begin{split}
   {\rm{H=}} \frac{1}{{\sqrt 2 }}\left[ {\begin{array}{*{20}{c}}
1&1\\1&{ - 1}\end{array}} \right], 
 {\rm{ X= }}\left[ {\begin{array}{*{20}{c}}0&1\\1&0\end{array}} \right],
{\rm{ CNOT= }}\left[ {\begin{array}{*{20}{c}}
1&0&0&0\\0&1&0&0\\0&0&0&1\\0&0&1&0\end{array}} \right], 
\\
 {\rm{ Toffoli= }}\left[ {\begin{array}{*{20}{c}}
{\begin{array}{*{20}{c}}
1&0&0&0\\
0&1&0&0\\
0&0&1&0\\
0&0&0&1
\end{array}}&{\begin{array}{*{20}{c}}
0&0&0&0\\
0&0&0&0\\
0&0&0&0\\
0&0&0&0
\end{array}}\\
{\begin{array}{*{20}{c}}
0&0&0&0\\
0&0&0&0\\
0&0&0&0\\
0&0&0&0
\end{array}}&{\begin{array}{*{20}{c}}
1&0&0&0\\
0&1&0&0\\
0&0&0&1\\
0&0&1&0
\end{array}}
\end{array}} \right],
 {\rm{ CSWAP= }}\left[ {\begin{array}{*{20}{c}}
{\begin{array}{*{20}{c}}
1&0&0&0\\
0&1&0&0\\
0&0&1&0\\
0&0&0&1
\end{array}}&{\begin{array}{*{20}{c}}
0&0&0&0\\
0&0&0&0\\
0&0&0&0\\
0&0&0&0
\end{array}}\\
{\begin{array}{*{20}{c}}
0&0&0&0\\
0&0&0&0\\
0&0&0&0\\
0&0&0&0
\end{array}}&{\begin{array}{*{20}{c}}
1&0&0&0\\
0&0&1&0\\
0&1&0&0\\
0&0&0&1
\end{array}}
\end{array}} \right].
\end{split}
\end{equation}

In quantum image processing, the complexity of the quantum circuit depends on the number of basic logic gate used. Single-qubit and two-qubit gates are basic quantum logic gates, and they can operate on arbitrary qubits. Therefore, we calculate the complexity of the circuit in terms of the number of single-qubit and two-qubit gates, which  is also called quantum cost. The quantum cost of a NOT gate, a reset gate or a CNOT gate is all 1. A Toffoli gate can be composed of 5 two-qubit gates, so its quantum cost is 5 \cite{Li2020,Yuan2020}.

\subsection{NEQR model}

Classical images contain position information and gray-scale value information, and the NEQR model uses three entangled sequences of qubits to represent these two kinds of information. Suppose a classical image's size is $2^n\times 2^n$, and the grayscale range is $[0, 2^q-1]$. Therefore, two $n$-length qubit sequences are needed to represent the position information, and one $q$-length qubit sequences are needed to represent the gray-scale value information. The NEQR expression can be written as  Eq. \ref{Eq1} \cite{Zhang2013}.

{\begin{equation}\label{Eq1}
 \lvert {I_{YX}} \rangle    = \frac{1}{{{2^n}}}\sum\limits_{Y = 0}^{{2^n} - 1} {\sum\limits_{X = 0}^{{2^n} - 1} {\lvert {{C_{YX}}} \rangle  \otimes \lvert Y \rangle \lvert X \rangle } }  = \frac{1}{{{2^n}}}\sum\limits_{YX = 0}^{{2^{2n}} - 1} {\mathop  \otimes \limits_{k = 0}^{q - 1}\lvert {C^K_{YX}} \rangle \mathop  \otimes \limits_{}^{} \lvert {YX} \rangle  }
\end{equation}
where  $\lvert {{C_{YX}}} \rangle  = \lvert {C_{YX}^{q - 1},C_{YX}^{q - 2},{ \cdots ^{}}C_{YX}^{1}C_{YX}^{0}} \rangle$ represents the quantum image gray-scale values, $C_{YX}^k \in \left\{ {0,1} \right\}$, $k = q - 1,q - 2, \cdots ,0$. $  \lvert {YX} \rangle  = \lvert Y \rangle \lvert X\rangle  = \lvert {{Y_{n - 1}},{Y_{n - 2}}, \cdots {Y_0}}\rangle \lvert {{X_{n - 1}},{X_{n - 2}}, \cdots {X_0}} \rangle $ represents the 
position of the pixel in a quantum image, ${Y_t},{X_t} \in \left\{ {0,1} \right\}$.}

\begin{figure}
 \centering
    \includegraphics[width=3cm]{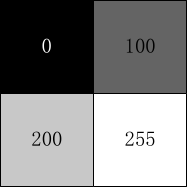}
    \caption{An example of a 2×2 image.}
    \label{An example of 2×2 image}
\end{figure}

Fig. \ref{An example of 2×2 image} shows an example of a grayscale image of size 2×2, and the corresponding NEQR expression of which is given as follows.
\begin{equation}
\begin{split}
     \begin{array}{l}
\lvert I \rangle  = \frac{1}{2}\left( {\lvert 0 \rangle \lvert{00} \rangle  + \lvert {100} \rangle \lvert {01} \rangle  + \lvert {200} \rangle \lvert {10} \rangle  + \lvert{255} \rangle \lvert {11} \rangle } \right)
\end{array} 
\\
\begin{array}{l}
=\frac{1}{2}\left( \begin{array}{l}
\lvert {00000000} \rangle \lvert{00} \rangle  + \lvert {01100100} \rangle \lvert {01} \rangle \\
 + \lvert {11001000}\rangle \lvert{10} \rangle  + \lvert {11111111} \rangle \lvert {11} \rangle 
\end{array} \right).
\end{array}
\end{split}
\end{equation}

\subsection{The classical local adaptive threshold for image segmentation}
\begin{figure}
    \centering
    \subfigure[]{\includegraphics[width=6cm]{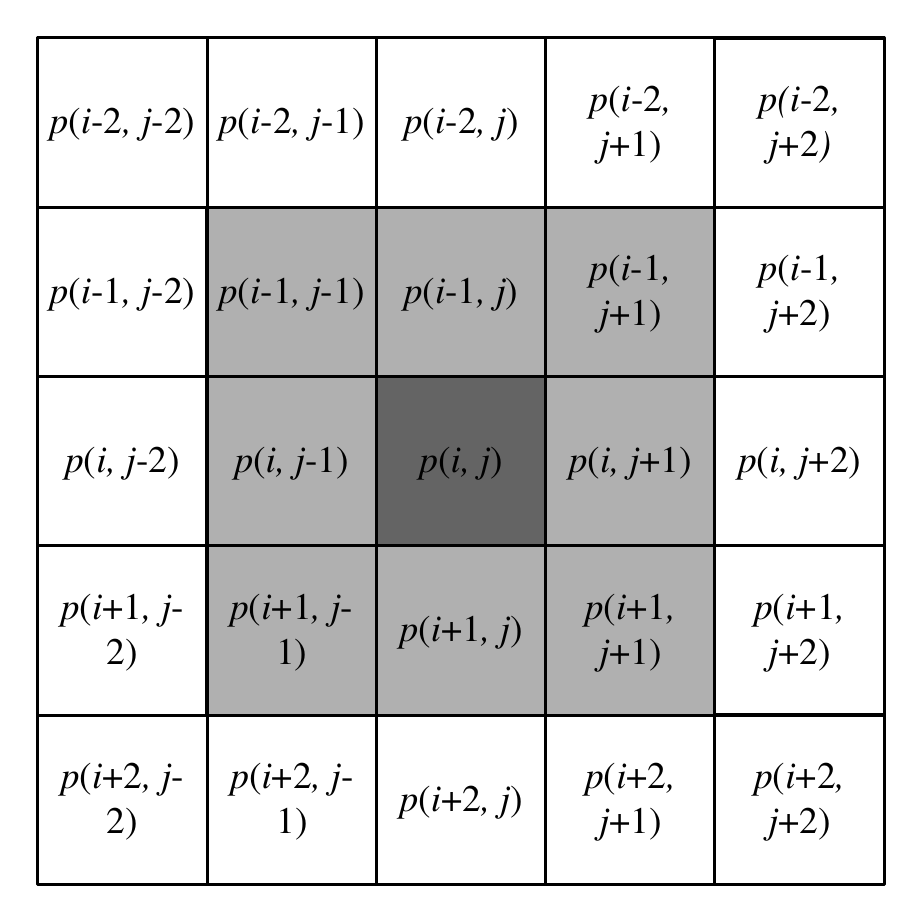}}
    \subfigure[]{\includegraphics[width=6cm]{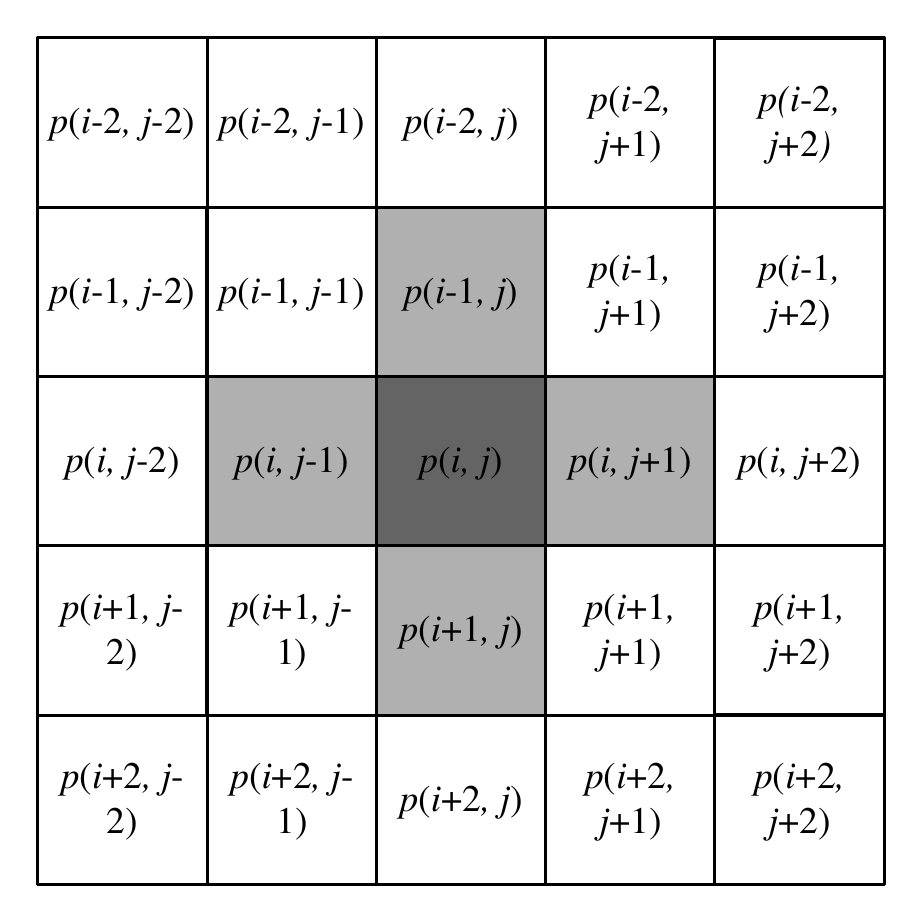}}
    \subfigure[]{\includegraphics[width=6cm]{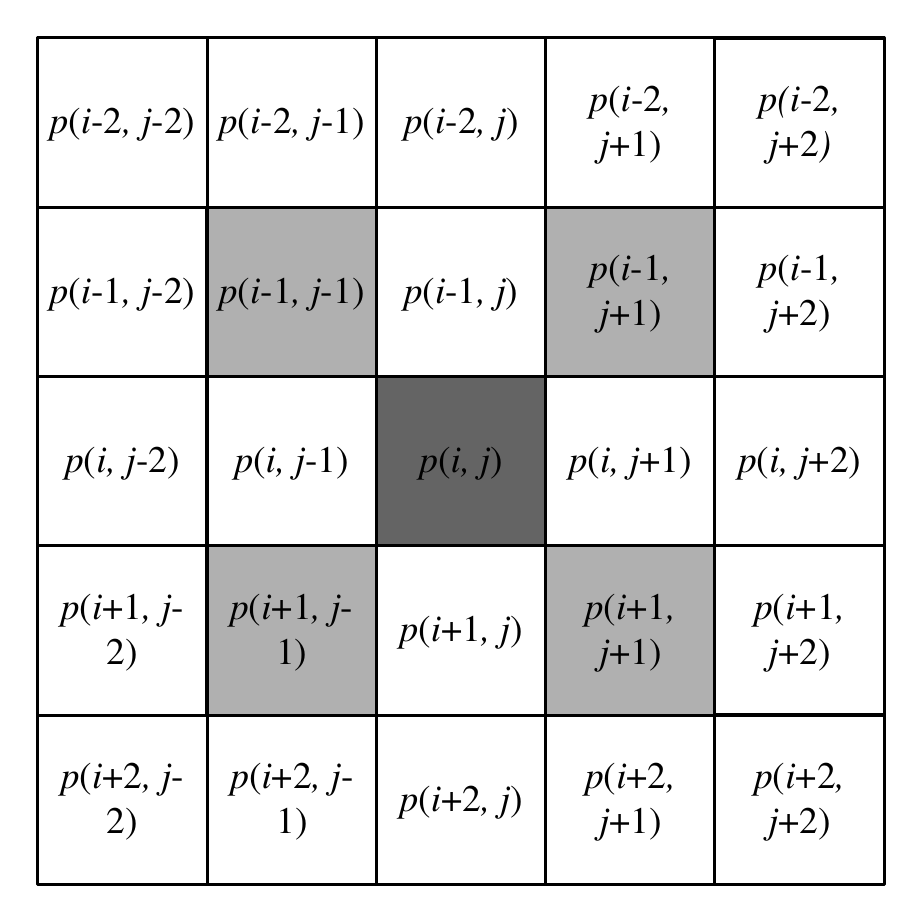}}
    \caption{ The schematic diagram of the neighborhood window.} 
    \label{The schematic diagram of the neighborhood window}
\end{figure}

Threshold-based image segmentation works by setting all pixels with intensities above a threshold as one value and all remaining pixels as another value. Traditional threshold operations use a global threshold for all pixels, which is inappropriate for images with uneven illumination. The local adaptive threshold operation can dynamically change the threshold according to the gray-scale values of the local neighborhood pixels, which can adapt to the illumination changes in the image. Adaptive threshold usually takes a gray-scale or color image as input, and in the simplest implementation, outputs a binary image. For each pixel in the image, it is set to 0 if the pixel value is below the threshold, otherwise set to 1. Using the median value of the neighborhood pixel intensities as a threshold is one of the most commonly used local adaptive threshold methods. In order to better find the threshold, the neighborhood window must be large enough to cover enough foreground and background pixels, but this violates the principle of local similarity, so it is necessary to subtract a  constant  $Z$   from the found median value as a new threshold. Using different neighborhood windows for different situations will achieve better results, and the commonly used windows are square, cross-shaped, diagonal, etc., 
as shown in Fig. \ref{The schematic diagram of the neighborhood window}. A square window is suitable for images containing objects with long outer contour lines, a cross-shaped window is suitable for images with more spikes and angles, and a diagonal window is suitable for images containing radial objects. Choosing the right neighborhood window can preserve the details in the image better. In this paper, we introduce the design of quantum circuits using a cross-shaped neighborhood window as an example. 

The gray-scale value of the original image is represented by $f(x,y)$, and $T$ is the threshold, which is obtained by subtracting the  constant  $Z$   from the median of the neighborhood window. The corresponding gray-scale value of the segmented image is represented by $b(x,y)$. The specific expression is described as follows:
\begin{equation}
    b(x,y) = \left\{ \begin{array}{l}
0, 0 \le f(x,y) < T\\
1,T \le f(x,y) \le 255
\end{array} \right.
\end{equation}
where $T= median-Z$, $Z$ is a constant we set.

\section{Quantum image segmentation algorithm and its quantum circuit implementation}\label{sec3}

In this section, several quantum operations are first introduced, and then a quantum image segmentation algorithm based on local adaptive thresholding is proposed. In addition,  the corresponding quantum circuit units are designd according to the algorithm steps, and finally a complete quantum segmentation circuit is given. We only mark the output information that will continue to be used, other information we do not need to consider and will be used as auxiliary qubits for the next operation.

\subsection{The quantum operations}

\begin{enumerate}[(1)]

\item Quantum comparator

The quantum comparator (QC) can compare the numerical magnitude relationship of two n-qubit sequences $ a $ and $ b$, where $a=\lvert a_{n-1}, a_{n-2}, \dots, a_{0} \rangle$, and $b=\lvert b_{n-1}, b_{n-2}, \dots, b_{0} \rangle $. The comparator takes two qubit sequences to be compared as input, and through a series of quantum logic gate operations, outputs the original two n-qubit sequences and their magnitude relationship result $y$. According to the comparison idea of quantum bit string comparator (QBSC) \cite{Oliveira2007}, the quantum comparator in this paper is also compared bit by bit. But in this paper, we only need to know the two relations $a\ge b$ or $a<b$, so we only need to use one auxiliary qubit to store the comparison result of each bit. As shown in Fig. \ref{The implementation circuit of the quantum comparator}, we implement the quantum comparator by using a combination of CNOT gates and Toffoli gates. We compare $a$ and $b$ bit by bit in order from low to high. There are four cases of $a_0b_0$, namely 00, 01, 10, and 11. Among them, 01 indicates $a<b$, and the other three cases indicate $a\ge b$. Therefore, only one Toffoli gate is needed to compare the numerical relationship between $a_0$ and $b_0$. 
If $a_0b_0=01$, then the auxiliary qubit $h_1$ is set to 1. Otherwise, it remains unchanged. $a_1b_1$ also has four cases of 00, 01, 10, and 11. We use the same processing method as $a_0b_0$. At this time, $h_2$ is set to 1 or unchanged. But when we synthesize the two comparison results, $a_1b_1a_0b_0=1001$ needs to be considered separately, so we use 2 Toffoli gates and one CNOT gate to process the two comparison results, which also requires an auxiliary qubit $h_0$ to store the second comparison result. After the two qubits comparisons, we use two reset operations \cite{Shende2005} to set the auxiliary qubits $h_2h_0$ to 00, so that we can reuse the two auxiliary qubits. Therefore, according to this comparison method, we can compare two n-qubit sequences, and if $a\ge b$, then $y=0$; if $a<b$, then  $y=1$. 

\begin{figure}
    \centering
    \includegraphics[width=12cm]{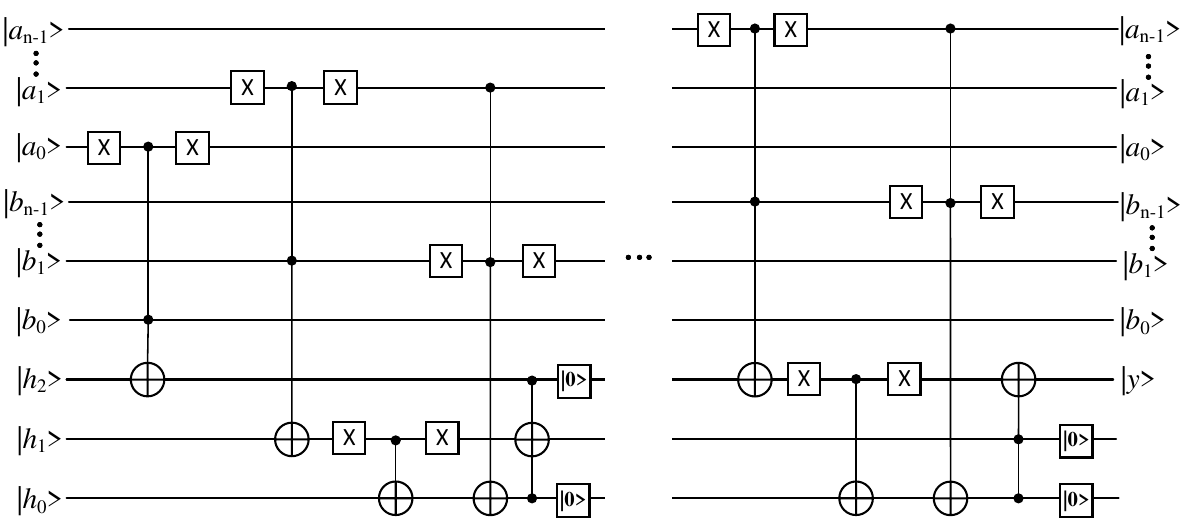}
    \caption{The implementation circuit of the quantum comparator (QC).}
    \label{The implementation circuit of the quantum comparator}
\end{figure}

\item Quantum subtractor

A quantum subtractor(QS) can perform subtraction operations on two binary qubits sequences  $a=\lvert a_{n-1}, a_{n-2}, \dots, a_{0} \rangle$ and $b=\lvert b_{n-1}, b_{n-2}, \dots, b_{0} \rangle $. As shown in Fig. \ref{The implementation circuit of the quantum subtractor}, the quantum subtractor takes two binary qubit sequences $a$ and $b$ as input and the result of $a-b$ as the output, where $a \ge b$. According to the idea of the ripple carry adder \cite{Thapliyal2013}, the quantum subtractor in this paper also operates bit by bit in the order from low to high. There are four cases of $a_0-b_0$: $0-0$, $0-1$, $1-0$, $1-1$. Among them, if $a_0-b_0$ is $0-1$, then $a_0$ needs to borrow from $a_1$. At this time, we use the auxiliary qubit $h_1$ to store the borrow information, and the output result of $a_0-b_0$ is 0. If $a_0-b_0$ is $0-0$ or $1-1$, then the output result is also 0. If $a_0-b_0$ is $1-0$, then the output result is 1. Therefore, if $b_0=1$, then we XOR $a_0$ to get the result of $a_0-b_0$. To save qubits, we use the reset operation and CNOT gate to transfer the borrow information of $a_0$ from $h_1$ to $b_0$, so that $h_1$ can be used in subsequent operations.  When calculating $a_1-b_1$, if $a_1-b_1=0-1$, then $a_1$ needs to borrow from $a_2$, and we store the borrow information in $h_1$; if $b_1=1$ or $a_1=0$, and $a_0$ borrows from $a_1$, then $a_1$ also need to borrow from $a_2$.  Finally, we aggregate the overall borrow information into $b_0$. The calculation method of $a_1-b_1$ is the same as that of $a_0-b_0$, except that we need to consider whether $a_0$ has borrowed from $a_1$. If $a_0$ has borrowed, then we need to perform an XOR operation on the result again. After these operations, we use the reset operation  to reset the auxiliary qubits so that we can continue to use them. The operation from $a_2-b_2$ to $a_{n-2}-b_{n-2}$ is the same as $a_1-b_1$, so we just need to repeat the operation to get the results. When calculating $a_{n-1}-b_{n-1}$, because $a>b$, $a_{n-1}$ will not borrow from higher qubits, we only need to consider whether $a_{n-2}$ borrows from $a_{n-1}$. At this time, in the four cases of $0-0$, $0-1$, $1-0$, $1-1$, if $b_{n-1}=1$, then the result of $a_{n-1}-b_{n-1}$ is the XOR of $a_{n-1}$, if $b_{n-1}=0$, then the result of $a_{n-1}-b_{n-1}$ is $a_{n-1}$. If $a_{n-2}$ borrows from $a_{n-1}$, then we  need to XOR the result  again. At this point, the results of $a-b$ can be completely output from the quantum subtractor.

\begin{figure}
    \centering
    \includegraphics[width=12cm]{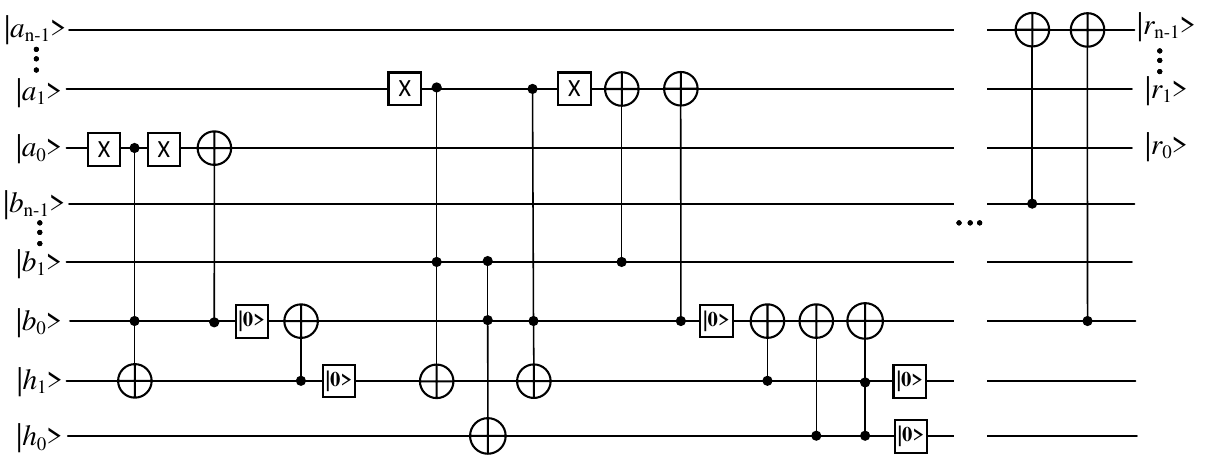}
    \caption{The implementation  circuit of the quantum subtractor (QS).}
    \label{The implementation circuit of the quantum subtractor}
\end{figure}

\item Quantum image cyclic shift transformation

The quantum image cyclic shift transformation (CT) operation (X shift and Y shift) \cite{Le2011S} can shift the positions of all pixels in an image , so that each pixel in the image can obtain its neighborhood pixels at the same time. For example, if we move the image  one unit down (CT$_y-$), the pixels in the image change from $S(X,Y)$ to $S(X,Y-1)$. The unitary operation for cyclic shift ransformation of a $2^n\times2^n$ NEQR image is shown as follows.

\begin{equation}
\mathrm{CT}(X \pm)\lvert I\rangle=\frac{1}{2^{n}} \sum_{Y=0}^{2^{n}-1} \sum_{X=0}^{2^{n}-1}\lvert C T_{Y X^{\prime}} \rangle \lvert Y\rangle \lvert (X \pm 1) \bmod 2^{n}\rangle 
\end{equation}
\begin{equation}
\mathrm{CT}(Y \pm)\lvert I\rangle=\frac{1}{2^{n}} \sum_{Y=0}^{2^{n}-1} \sum_{X=0}^{2^{n}-1}\lvert C T_{Y^{\prime} X} \rangle \lvert (Y \pm 1) \bmod 2^{n}\rangle \lvert X \rangle 
\end{equation}
where $X^{\prime}=(X \mp  1) \bmod 2^{n}$, $Y^{\prime}=(Y \mp  1) \bmod 2^{n}$, $\mathrm{CT}_{(X+)} \, \&\, \mathrm{CT}_{(Y+)}=\left[ {\begin{array}{*{20}{c}}
0&1\\
{I_2^n - 1}&0
\end{array}} \right]$, $\mathrm{CT}_{(X-)} \,\&\, \mathrm{CT}_{(Y-)}=\left[ {\begin{array}{*{20}{c}}
0&{I_2^n - 1}\\
1&0
\end{array}} \right]$. Fig. \ref{Quantum circuit for cyclic shift transformation} shows a schematic diagram of the X-axis cyclic shift ransformation.

From the above analysis, we can know that when an image is subjected to CT operation, the edge of the image will be cyclically shifted to the other side, as shown in Fig. \ref{The schematic diagram of cyclic shift}.

\begin{figure}
    \centering
    \subfigure[]{\includegraphics[width=5.5cm]{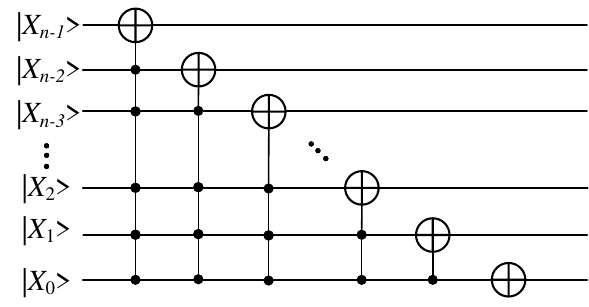}}
    \subfigure[]{\includegraphics[width=5.5cm]{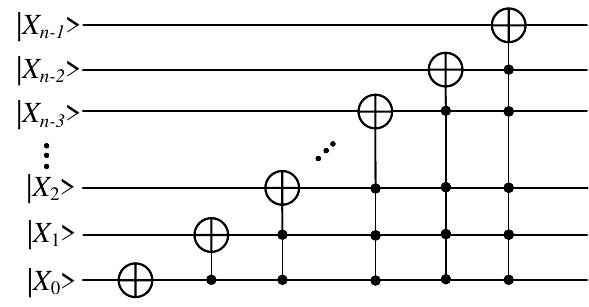}}
    \caption{Quantum circuit for cyclic shift transformation (CT): a CT($X+$), b CT($X-$).}
    \label{Quantum circuit for cyclic shift transformation}
\end{figure}

\begin{figure}
    \centering
   \subfigure[]{\includegraphics[width=3.5cm]{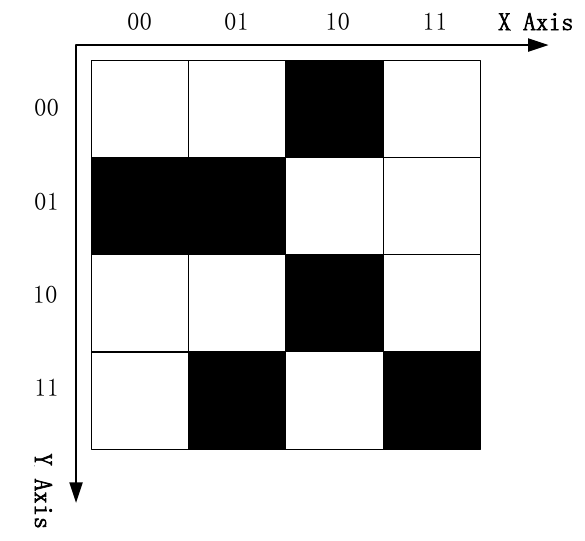}} 
   \subfigure[]{\includegraphics[width=3.5cm]{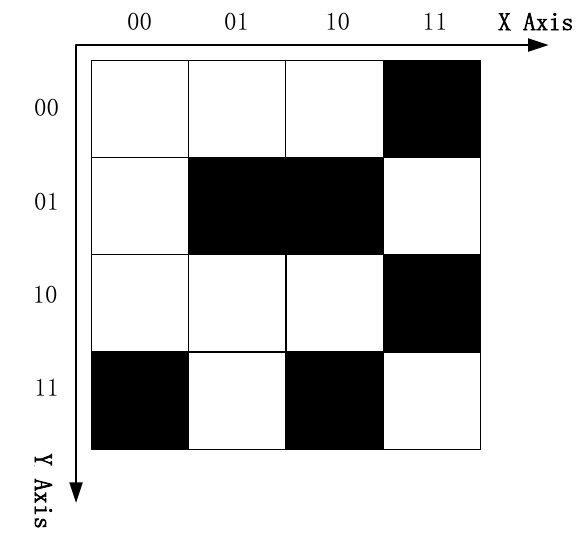}} 
   \subfigure[]{\includegraphics[width=3.5cm]{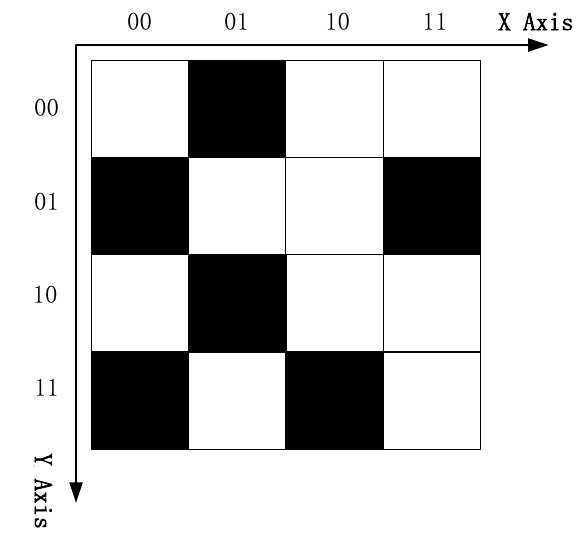}} 
    \caption{The schematic diagram of cyclic shift: a An example of original image of size $4\times4$, b the image transformed after CT($X+$) operation, c the image transformed after CT($X-$) operation.}
    \label{The schematic diagram of cyclic shift}
\end{figure}

\end{enumerate}

\subsection{The quantum segmentation algorithm}
\begin{figure}
    \centering
    \includegraphics[width=11cm]{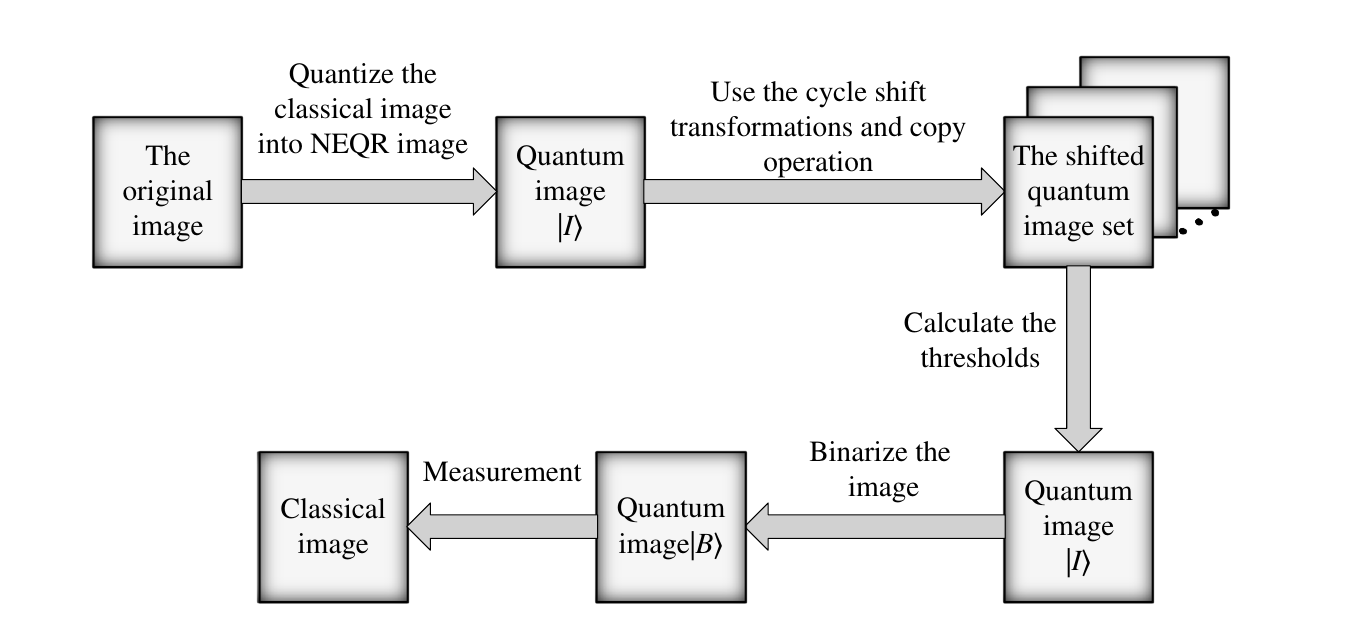}
    \caption{The workflow of our proposed algorithm.}
    \label{The workflow of our proposed algorithm}
\end{figure}

The image segmentation  based on threshold is a simple and fast segmentation algorithm, but the use of global threshold cannot deal with the uneven illumination in the image. Therefore, in this paper, a quantum image segmentation algorithm based on local adaptive threshold is proposed.  First the original image is cyclically shifted according to the neighborhood window to prepare a quantum image set, so we can process all pixels at the same time. Then, we use the median of the neighborhood window pixels minus the  constant  $Z$   as the threshold to segment the center pixel, and  the pixels greater than or equal to the threshold are converted to 1,   the  remain are converted to 0. Finally, we can get a binary image. The cross-shaped neighborhood mask  is used as an instance to show the procedure of image segmentation in detail. According to this circuit design method, it can be extended to any neighborhood window mask. The specific workflow is shown in Fig. \ref{The workflow of our proposed algorithm}.

\subsection{Quantum circuit implementation}

\textbf{Step 1 } Because quantum image cannot be directly obtained at present, we need to quantize the classical image into NEQR image. $(2n+q)$ qubits are needed to store a  $2^n\times2^n$ classical image with   $q$ gray-scale levels. Furthermore, additional $4q$ qubits are needed to store the gray-scale value of the 4-neighborhood pixel window, and another $q$ qubits are also needed to replicate the gray-scale values of the original image. The quantum state expression can be represented as
\begin{equation}
   \begin{array}{l}
{\lvert 0 \rangle ^{ \otimes 5q}} \otimes \lvert {{I_{YX}}} \rangle  = \frac{1}{{{2^n}}}\sum\limits_{Y = 0}^{{2^n} - 1} {\sum\limits_{X = 0}^{{2^n} - 1} {{{\lvert 0 \rangle }^{ \otimes 5q}}\lvert {{C_{YX}}} \rangle } } \lvert Y \rangle \lvert X \rangle \\
\begin{array}{*{20}{c}}
{}&{}&{}&{}&{}&{}{}&{}&{\begin{array}{*{20}{c}}
{}&{}&{}&{}&{}&{}{}&{}&{}& = 
\end{array}}
\end{array}\frac{1}{{{2^n}}}\sum\limits_{Y = 0}^{{2^n} - 1} {\sum\limits_{X = 0}^{{2^n} - 1} {{{\lvert 0 \rangle }^{ \otimes q}}} }  {\lvert 0 \rangle ^{ \otimes q}}{\lvert 0 \rangle ^{ \otimes q}} {\lvert 0 \rangle ^{ \otimes q}}{\lvert 0 \rangle ^{ \otimes q}} \lvert {{C_{YX}}} \rangle \lvert Y \rangle \lvert X \rangle 
\end{array} 
\end{equation}

\textbf{Step 2} According to the neighborhood window, we need to perform a cyclic shift transformation on the original image $\lvert{I_{YX}}\rangle$ to prepare a quantum image set, so that we can process all  pixels in the image simultaneously. First, we translate the position of the original image by one unit up, down, left, and right, respectively, as shown in Table \ref{tab1}. A cyclic shift in each direction results in a new NEQR image, so we have a total of 5 NEQR images. Each pixel at the same location in these 5 NEQR images is exactly the 4-neighborhood window pixel in the original image. Since the gray-scale values of the NEQR image are superimposed, when we operate on the neighborhood window of a pixel in the original image, we can operate on the neighborhood windows of all pixels at the same time. In addition, since we still need to use the original image after sorting, but the five images have been mixed together during the sorting process, we need to use the CNOT gate to copy the original image into an additional 3 qubits for later  binarization operation.
 The shifted quantum image sets can be encoded as follows.

  \begin{equation} 
 \begin{array}{l}
\frac{1}{{{2^n}}}\sum\limits_{Y = 0}^{{2^n} - 1} {\sum\limits_{X = 0}^{{2^n} - 1} {\lvert {{C_{Y + 1X}}} \rangle } }  \otimes \lvert {{C_{YX + 1}}} \rangle  \otimes \lvert {{C_{Y - 1X}}} \rangle \\
 \otimes \lvert {{C_{YX - 1}}} \rangle  \otimes \lvert {{C_{YX}}} \rangle  \otimes \lvert {{C_{YX}}} \rangle  \otimes \lvert Y \rangle \lvert X \rangle 
\end{array}
\end{equation}
    
 \begin{table}
\tbl{Computation prepared algorithm for shifting the image.}
{\begin{tabular}{@{}c@{}} \toprule
1. Input: the original NEQR image ${I_{YX}}$,$ \lvert {{I_{YX}}} \rangle  = \frac{1}{{{2^n}}}\sum\limits_{Y = 0}^{{2^n} - 1} {\sum\limits_{X = 0}^{{2^n} - 1} {\lvert {{C_{YX}}} \rangle } } \lvert Y \rangle \lvert X \rangle $\\  
2. Shift ${I_{YX}}$one unit upward, then$\lvert  {{I_{Y + 1X}}} \rangle  = \frac{1}{{{2^n}}}\!\sum\limits_{Y = 0}^{{2^n} - 1} \!{\sum\limits_{X = 0}^{{2^n} - 1} {\lvert  {{C_{Y + 1X}}} \rangle } } \lvert  Y \rangle \lvert  X \rangle $\\
3. Shift${I_{YX}}$one unit downward, then$ \lvert {{I_{Y - 1X}}} \rangle  =  \frac{1}{{{2^n}}}\sum\limits_{Y = 0}^{{2^n} - 1} {\sum\limits_{X = 0}^{{2^n} - 1} {\lvert  {{C_{Y - 1X}}} \rangle } } \lvert  Y \rangle \lvert  X \rangle $\\
4. Shift${I_{YX}}$ one unit leftward,  then then$\lvert {{I_{YX + 1}}} \rangle  =  \frac{1}{{{2^n}}}\sum\limits_{Y = 0}^{{2^n} - 1} {\sum\limits_{X = 0}^{{2^n} - 1} {\lvert {{C_{YX + 1}}} \rangle } } \lvert Y \rangle \lvert X \rangle $\\

5. Shift${I_{YX}}$one unit rightward,  then $\lvert {{I_{YX - 1}}} \rangle  =  \frac{1}{{{2^n}}}\sum\limits_{Y = 0}^{{2^n} - 1} {\sum\limits_{X = 0}^{{2^n} - 1} {\lvert {{C_{YX - 1}}} \rangle } } \lvert Y \rangle \lvert X \rangle$\\
\botrule
\end{tabular}\label{tab1} }
\end{table}

\begin{figure}
    \centering
    \includegraphics[width=12cm]{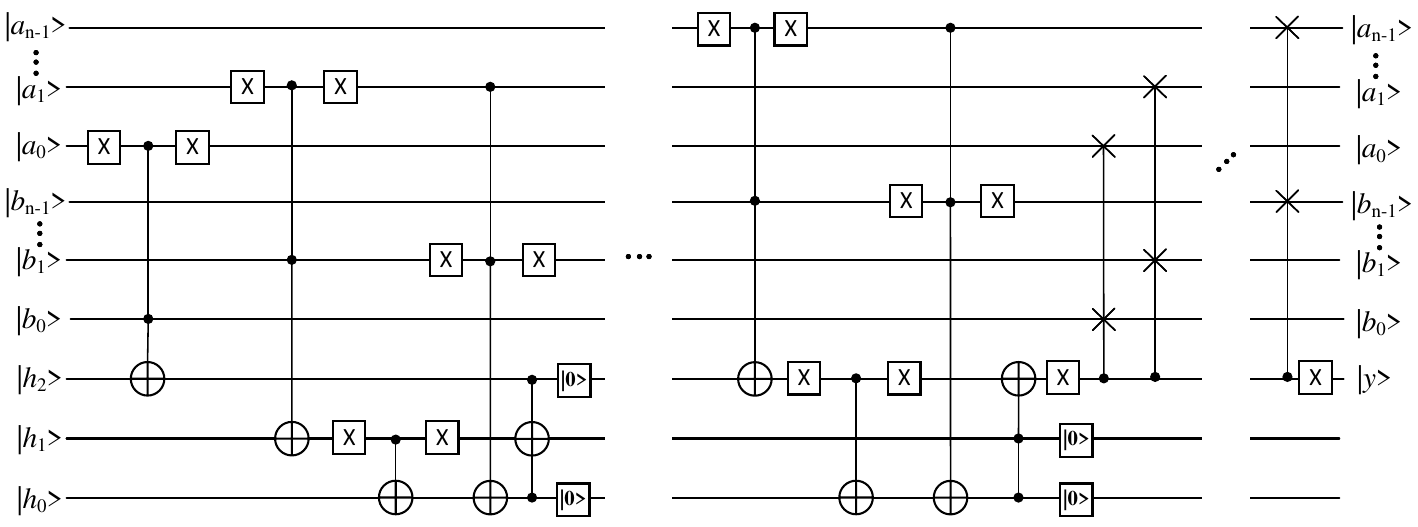}
    \caption{The implementation circuit of QCS.}
    \label{The implementation circuit of QCS}
\end{figure}

\begin{figure}
    \centering
    \includegraphics[width=9cm]{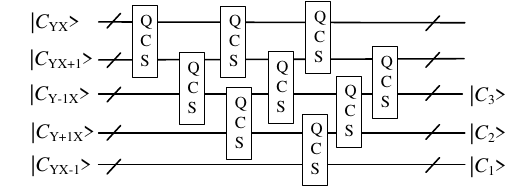}
    \caption{The implementation circuit for calculating the median.}
    \label{The implementation circuit for calculating median}
\end{figure}

\begin{figure}
    \centering
    \includegraphics[width=4cm]{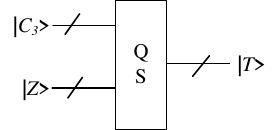}
    \caption{The implementation circuit for calculating the threshold $T$.}
    \label{The implementation circuit for calculating the threshold T}
\end{figure}

\textbf{Step 3 } 
In order to facilitate the sorting of the neighborhood window pixels, we construct the QCS according to the QC, as shown in Fig. \ref{The implementation circuit of QCS}. When the output of the QC is $y=0$, we use the CSWAP gates to swap the positions of $a$ and $b$ and let the larger value be the input of the next comparator, which can find the largest number in the neighborhood window. We compare the central pixel of the four-neighborhood window with other pixels in turn. After each comparison, we output the larger number from the position of $b$ to the next QCS and then compare it with another pixel, and finally find the largest number $C_1$. In this way, we can find three larger numbers $C_1C_2C_3$ after three rounds of comparison. At this point, $C_3$ is the median we are looking for, and the specific operation circuit is shown in Fig. \ref{The implementation circuit for calculating median}. In addition, to adjust the threshold, we need to use QS to subtract a  constant  $Z$   from the median found. $Z$ is used to adjust the threshold value, and we need to set it manually according to the actual segmentation results. So that we can get the final threshold $T$, as shown in Fig. \ref{The implementation circuit for calculating the threshold T}. Because we only need to keep $C_3$, the  constant  $Z$   can be set in other pixel neighborhood pixel positions, and only the corresponding position needs to be reset \cite{Shende2005}, which can save qubits. What's more, $Z$ is a definite value, it only needs to be initialized with the reset operation and the NOT gate.

\begin{figure}
    \centering
    \includegraphics[width=8cm]{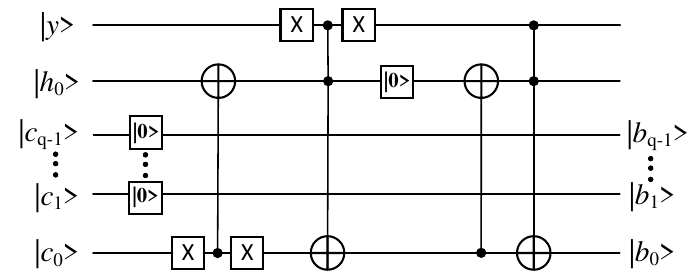}
    \caption{The implementation  circuit of quantum binarization (QB).}
    \label{The implementation circuit of binarization}
\end{figure}

\textbf{Step 4 } After getting the threshold $T$, we need to change the pixel greater than or equal to $T$ to 1, and change the other pixels to 0. In this way, we can clearly segment the object we want. In this step, we use the reset operation to set $c_{q-1}\dots c_1$ to $0\dots 0$. If QC outputs $y=0$ and $c_0=0$, then we use CNOT gate and Toffoli gate to set $c_0$ to 1. If QC outputs $y=1$ and $c_0=1$, then we use CNOT gate and Toffoli gate to set $c_0$ to 0. Finally, we can get a binary image $\lvert B_{YX} \rangle$, and the specific operation is shown in  Fig. \ref{The implementation circuit of binarization}. The result image $\lvert B_{YX} \rangle$ can be represented  as follows.

\begin{equation}
  \lvert {B_{YX}} \rangle     = \frac{1}{{{2^n}}}\sum\limits_{Y = 0}^{{2^n} - 1} {\sum\limits_{X = 0}^{{2^n} - 1} {\lvert {{b_{YX}}} \rangle  \otimes \lvert Y \rangle \lvert X \rangle } }
\end{equation}

\begin{figure}
    \centering
    \includegraphics[width=12cm]{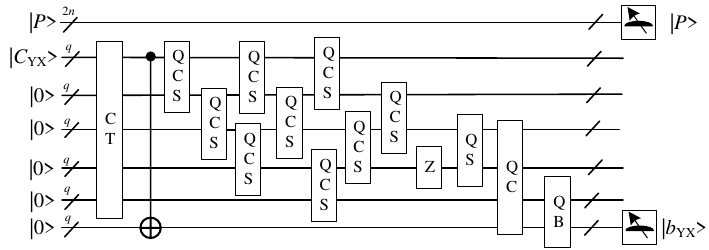}
    \caption{The complete quantum circuit of our quantum image segmentation algorithm.}
    \label{The complete quantum circuit of our quantum image segmentation algorithm}
\end{figure}

The complete quantum circuit of our quantum image segmentation algorithm is shown in Fig. \ref{The complete quantum circuit of our quantum image segmentation algorithm}.

\section{Circuit complexity and experiment analysis }\label{sec4}
\subsection{Circuit complexity  analysis}

 Taking a $2^n\times 2^n$ classical image with gray-scale $[0,q]$  as an example, we will discuss the  complexity of the circuit in 4 steps.

In  step 1, we need to prepare the classical image into a NEQR image, and it is known from Ref. \cite{Zhang2013} that the computational complexity of this step does not exceed O($qn2^{2n}$), because  each pixel will be operated one after another. However, our algorithm is to process quantum images, not classical images, so the complexity of this step is not considered in the study of quantum image processing algorithms \cite{Zhou2019,Yuan2020,Chetia2021}, so we consider the complexity of this step to be 0.

In step 2, we need to use CT operation to cycle shift the image, and the complexity of CT is O$(n^2)$ \cite{Wang2014,Le2011S}.  In addition, we also use $q$ CNOT gates to replicate the gray-scale value of the original image, the quantum cost of this is $q$.
So the complexity of this step is O$(n^2+q)$.

In step 3, we use 9 QCS and 1 QS to calculate the threshold $T$.   The QCS is composed of a QC and $q$ CSWAP gates.  One QC requires $3q-2$ Toffoli gates, $(q-1)$ CNOT gates, and $2(q-1)$ reset gates. So the quantum cost of a QC is $18q-13$. In addition, the quantum cost of a CSWAP gate is 3 \cite{Li2020}, and $q$ CSWAP gates are needed. So the quantum cost of a QCS is $21q-13$. A QS requires $4q-7$ Toffoli gates, $4q-4$ CNOT gates, and $3q-4$ reset gates. Therefore, the quantum cost of a QS is $27q-43$. As can be seen from the above analysis, the complexity of this step is O$(q)$.

In step 4, we perform a quantum binarization (QB) operation on the image. As shown in the Fig. \ref{The implementation circuit of binarization}, this circuit requires 2 Toffoli gates, 2 CNOT gates and $q-1$ reset gates. So the quantum cost of this step is $q+11$, and the complexity is O$(q)$.

In addition, in order to set the  constant  $Z$  , we also need $q$ reset gates and $q$ NOT gates,  so the quantum cost is $2q$, and the complexity is O$(q)$. Therefore,the complexity of the complete algorithm is O$(n^2+q)$. On classical computers, for images of size $2^n \times 2^n$, image segmentation need to be processed
individually for each pixel. So, the complexity of the classical segmentation algorithm is no less than O$(2^{2n})$ . Thus, our scheme achieves an exponential acceleration relative to the classical segmentation algorithm, so the real-time problem in classical algorithm can be solved well.

\subsection{Experiment analysis}

\begin{figure}
    \centering
    \includegraphics[width=6cm]{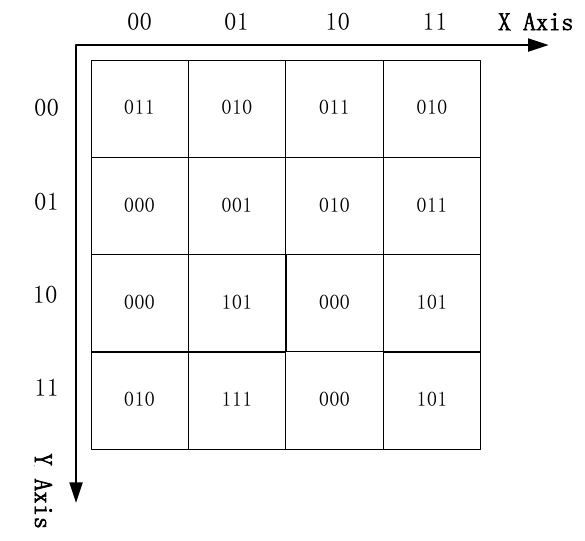}
    \caption{The schematic diagram of the  original image}
    \label{The original image}
\end{figure}

In order to verify the feasibility of the algorithm we proposed, the simulation experiment is conducted on the IBM Q. The IBM Q \cite{IBM} currently provides 24 quantum computers and 5 quantum simulators in the cloud. But due to certain factors, the quantum computer available to us has only 5 qubits, which is far from enough. Therefore, in this paper, we use a quantum simulator called 'ibmq\_qasm\_simulator' to conduct the experiment through the Qiskit extension package \cite{Qiskit}. This simulator has 32 qubits and can run all the basic logic gates included in our circuit. So in this paper, all quantum circuits have been simulated and verified on this simulator. But the actual image processing requires a large number of qubits, therefore, due to the limitation of the number of qubits, we selected a classic image with a size of $4\times4$ and a gray-scale value of [0,7]. The schematic diagram of the image is shown in Fig. \ref{The original image}.  Y and X represent the coordinates of the pixel, and  the gray-scale value information is marked in the figure in binary form. In addition,  noise will reduce the accuracy of segmentation, and how to reduce the noise by using quantum mechanisms is another interesting topic. Our algorithm only considers how to implement the classical algorithm using the quantum mechanism and does not consider the noise effect. Therefore, we assume that the experiment is performed without noise.  

\begin{figure}
    \centering
    \includegraphics[width=7.5cm]{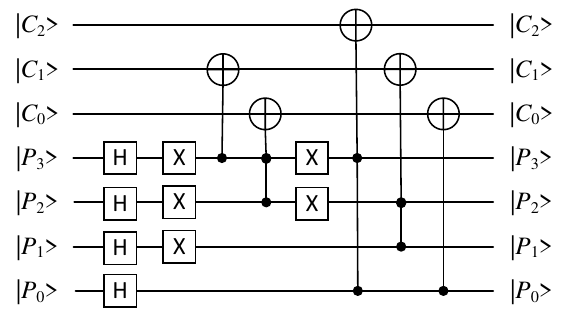}
    \caption{The implementation circuit of quantum image preparation.}
    \label{The implementation circuit of quantum image preparation}
\end{figure}

First, we need to prepare a NEQR image based on the gray-scale value information and position information of the classical image. Because the NEQR model makes full use of quantum entanglement and superposition, we only need four qubits to store $4 \times 4$ position information, and then we entangle the gray-scale value and position. So only three qubits are needed to store all the gray-scale values in the image. The quantum image preparation circuit is shown in the Fig. \ref{The implementation circuit of quantum image preparation}, where $C_2C_1C_0$ represent the gray-scale value information qubits, and $P_3P_2P_1P_0$ represent the position information qubits. Because on the IBM Q platform, the qubits are initialized to 0 by default, so we use H gates to make the position information qubits into a superposition state. The gray-scale value information is entangled with position by using basic quantum gates.     Besides, in the experiment we set the  constant  $Z=001$ randomly for the adjustment of the threshold to to verify the feasibility of the algorithm.  Because $Z$ is a definite value, it only needs to be initialized with the reset operations and the NOT gates.

\begin{figure}
    \centering
    \includegraphics[width=10cm]{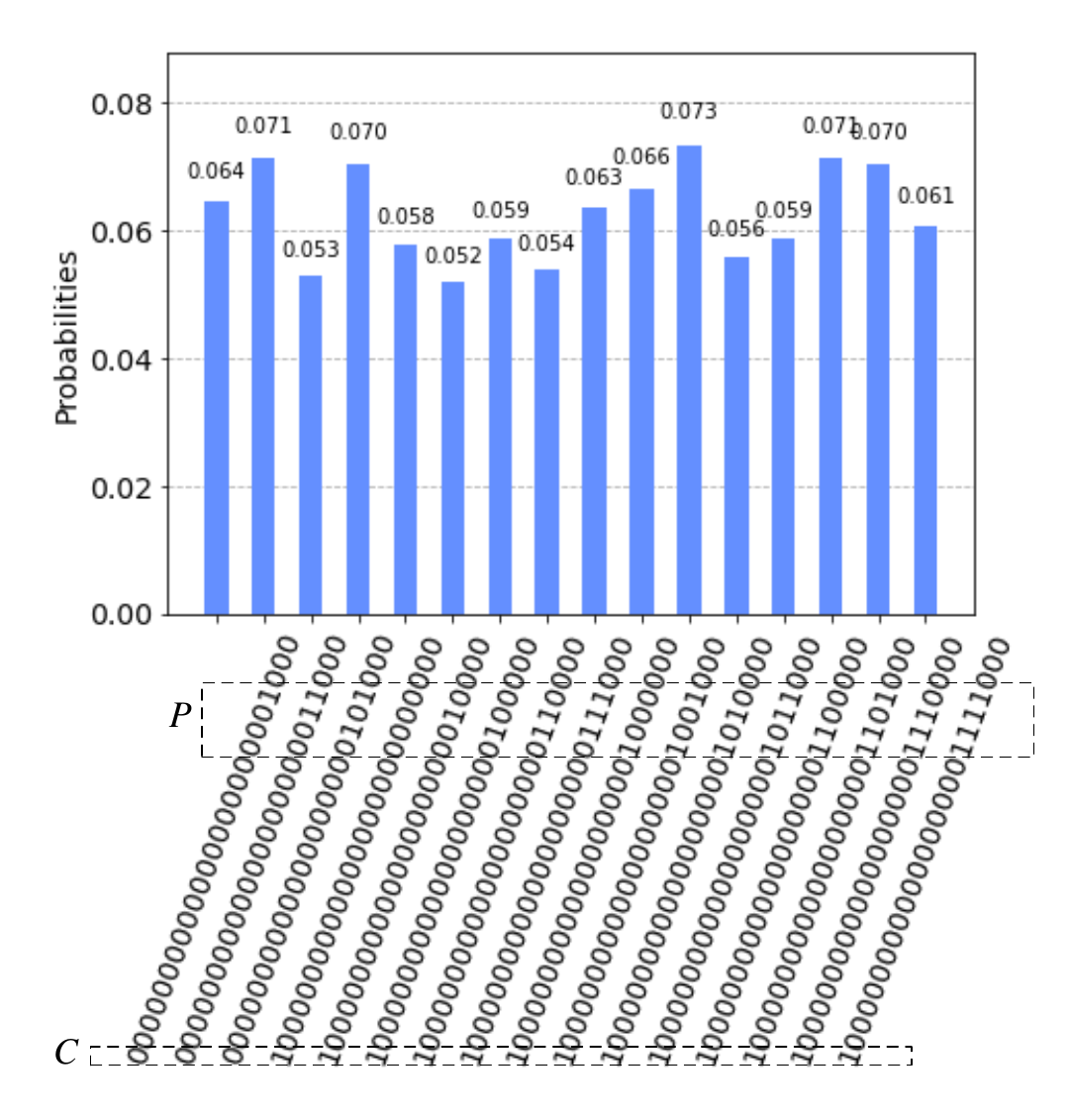}\caption{The probability histogram of the result image.}  
    \label{The Probability histogram of the result image}
\end{figure}

\begin{figure}
    \centering
    \includegraphics[width=6cm]{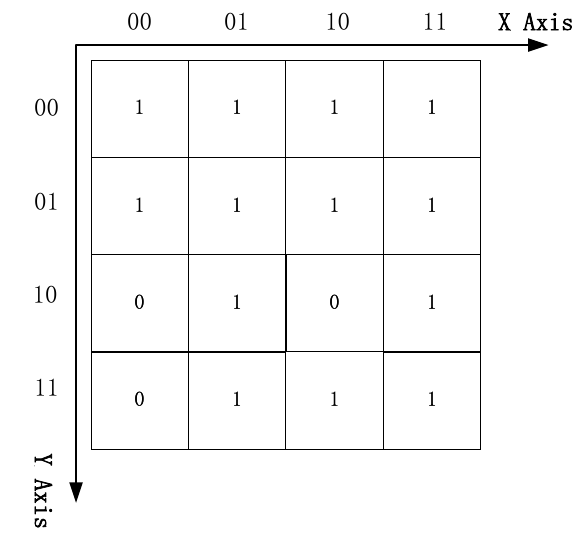}
    \caption{The schematic diagram of the result image.}
    \label{The schematic diagram of the result image}
\end{figure}

 \begin{figure}
    \centering
   \subfigure[]{\includegraphics[width=3.5cm]{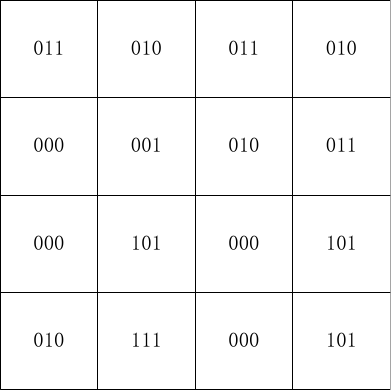}} 
  \subfigure[]{\includegraphics[width=3.5cm]{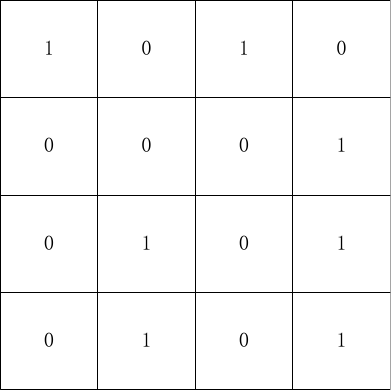}} 
  \subfigure[]{\includegraphics[width=3.5cm]{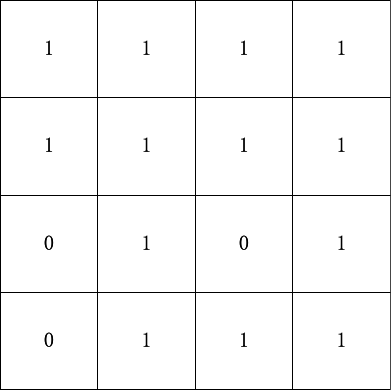}} \\
  \setcounter{subfigure}{0}\subfigure[]{\includegraphics[width=3.5cm]{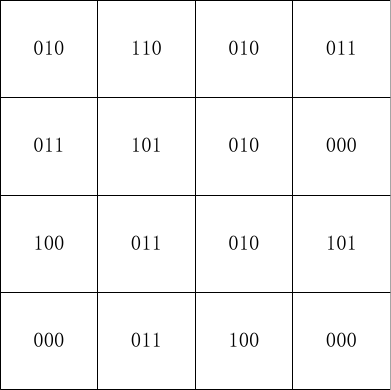}} 
  \subfigure[]{\includegraphics[width=3.5cm]{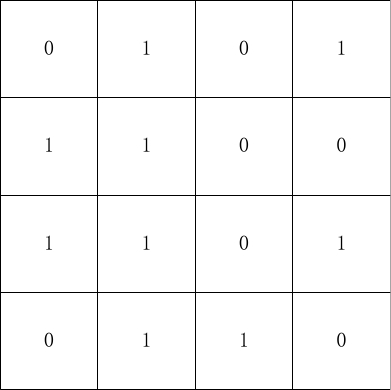}} 
  \subfigure[]{\includegraphics[width=3.5cm]{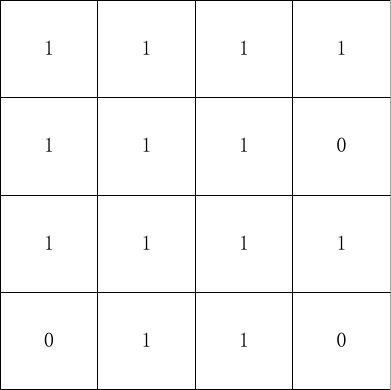}} \\
  \setcounter{subfigure}{0}\subfigure[]{\includegraphics[width=3.5cm]{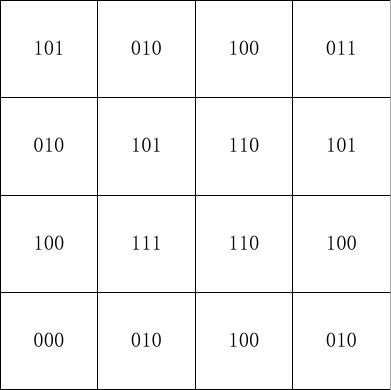}} 
  \subfigure[]{\includegraphics[width=3.5cm]{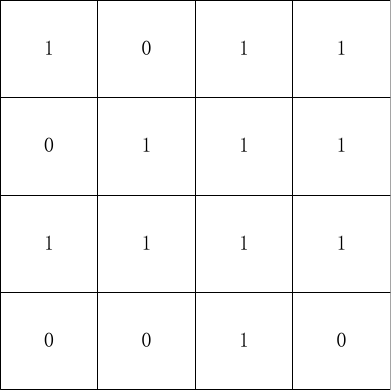}} 
  \subfigure[]{\includegraphics[width=3.5cm]{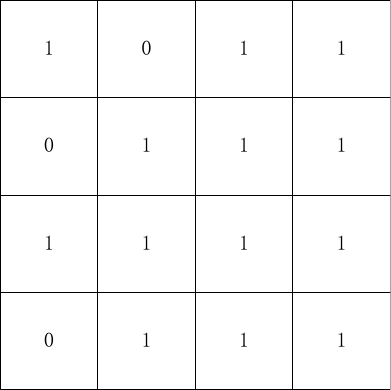}} 
    \caption{The comparison of different segmentation results: a  original images, b the image segmented by fixed threshold ($T=3$), c the image segmented by our algorithm.} 
    \label{The comparison of different segmentation results}
\end{figure}

 Fig. \ref{The Probability histogram of the result image} shows the probability histogram of the result quantum image, where the ordinate represents the probability of the measured qubits sequences, and the abscissa represents each pixel information and other auxiliary qubit information. Each qubit sequence is in order from top to bottom,  gray-scale value information C and the position information P of the result image are marked in the figure.   The remaining qubits are used to form a complete circuit, which has no effect on image representation, so to reduce the complexity of the measurement, we only measured the gray-scale value and position information qubits of the result image. A schematic diagram of the result image is shown in Fig. \ref{The schematic diagram of the result image}, and the position information and gray-scale value information have been marked in  it.
 In addition, We randomly selected three images as test images to compare our algorithm with the fixed-threshold algorithm, as shown in Fig. \ref{The comparison of different segmentation results}. For better comparison of segmentation results,   we use the mean square error (MSE) to evaluate our algorithm.
MSE can be used to determine the similarity of two images. When the MSE value is smaller, it means that the two images are more similar, which indicates that the segmentation algorithm will be more effective.  For two gray-scale images Q and R with size $2^n\times2^n$, MSE is defined as 

\begin{equation}
    MSE=\frac{1}{2^{2n}} \sum_{Y=0}^{2^{n}-1} \sum_{X=0}^{2^{n}-1}[Q(Y, X)-R(Y, X)]^{2}
\end{equation}
where Y and X represent the position information of the images. 

As shown in Tab. \ref{Tab2}, the MSE values of our algorithm are less than the fixed threshold segmentation, which indicates that our algorithm has a better segmentation effect. Since we use local adaptive threshold and find the threshold within the neighborhood window, this prevents the effect of uneven illumination on image segmentation. 

 \begin{table}[]
\tbl{The MSE comparison of different segmentation results.} {
\begin{tabular}{ccc}
\hline
\multirow{2}{*}{Input image} & \multicolumn{2}{c}{MSE}         \\ \cline{2-3} 
                             & Fixed threshold & Our algorithm \\ \hline
a                            &  7.0625          & 6.5625        \\
b                            &  14.2500          &  11.8750       \\
c                            &  11.8750              & 11.5000            \\ \hline
\end{tabular}}\label{Tab2}
\end{table}

\section{Conclusion}\label{sec5}

Since most of the current quantum image segmentation algorithms use global fixed thresholds and  require more quantum gates and qubits, whis is not applicable in this NISQ era. In this paper, a quantum segmentation algorithm based on local adaptive threshold for NEQR image is proposed, which can use quantum mechanism to simultaneously compute local thresholds for all pixels in a gray-scale image and quickly segment the image into a binary image.  Furthermore, we designed several quantum circuit units and  a complete quantum segmentation circuit  by using fewer qubits and quantum gates. Circuit complexity analysis and experiment on IBM Q demonstrate the superiority and feasibility of our algorithm.

However,  the local adaptive threshold segmentation algorithm only focuses  on the neighborhood window and does not consider the noise effect, which may lead to some noise in the result image, so our future work  is to investigate  quantum image noise reduction algorithms and apply them to quantum image segmentation.

\section*{Acknowledgments}

This work is supported by the National Natural Science Foundation of China (62071240),  and the Priority Academic Program Development of Jiangsu Higher Education Institutions (PAPD).

\end{document}